\documentstyle[prl,aps,epsf]{revtex}
\begin{document}

\title{Pair dispersion in synthetic fully developed turbulence}

\author{G. Boffetta, A. Celani}
\address{Dipartimento di Fisica Generale, Universit\`a di Torino,
         v. Pietro Giuria 1, 10125 Torino,}
\address{Istituto di Cosmogeofisica, c. Fiume 4, 10133 Torino}
\address{and Istituto Nazionale di Fisica della Materia, 
Unit\`a di Torino, Italy}

\author{A. Crisanti and A. Vulpiani}
\address{Dipartimento di Fisica, Universit\`a di Roma ``La Sapienza'',
         p.le Aldo Moro 2, 00185 Roma}
\address{and Istituto Nazionale di Fisica della Materia, 
Unit\`a di Roma I, Italy}

\date{\today}

\maketitle

\begin{abstract}
The Lagrangian statistics of relative dispersion in fully developed 
turbulence is numerically investigated.
A scaling range spanning many decades is achieved by generating a 
synthetic velocity field with prescribed Eulerian statistical
features. When the velocity field obeys Kolmogorov similarity,
the Lagrangian statistics is self similar too, and 
in agreement with Richardson's predictions.
For an intermittent velocity field the scaling laws for the 
Lagrangian statistics are found to depend
on Eulerian intermittency in agreement with a multifractal
description. As a consequence of the Kolmogorov law
the Richardson law for the variance of pair separation is not affected 
by intermittency corrections.
A new analysis method, based on fixed scale averages instead of usual
fixed time statistics, is shown to give much wider scaling 
range and should be preferred for the analysis of experimental data.
\end{abstract}



\section{Introduction}

Understanding the statistics of particle pairs dispersion 
in turbulent velocity fields is of great interest for both theoretical and 
practical implications. 
At variance with single particle dispersion, which depends mainly on
large scale, energy containing eddies, pair dispersion is driven
(at least at intermediate times) by velocity fluctuations at scales
comparable with the pair separation.
These small scale fluctuations
are thought to be independent on the particular large scale flow
\cite{Batchelor52}.
Since fully developed turbulence displays well known, non-trivial 
universal features in the Eulerian statistics of velocity differences 
\cite{MY75,Frisch95}, pair dispersion represents a starting point for 
the investigation
of the general problem of the relationship between Eulerian
and Lagrangian properties.

Since the pioneering work by Richardson \cite{Richardson26},
many efforts have been done to confirm experimentally \cite{MY75}
or numerically \cite{ZB94,EM96,FV98} his law. 
Nevertheless, the main obstacle to a deep 
investigation of relative dispersion in turbulence 
remains the lack of sufficient
statistics due to technical difficulties in laboratory experiments and to
the moderate inertial range reached in direct numerical simulations.

Moreover, also from an applicative point of view, a deep comprehension
of the relative dispersion mechanisms is of fundamental importance
for a correct modelization of small scale diffusion and mixing 
properties.

In this Paper we present a detailed investigation of the
statistics of relative dispersion, obtained by numerical simulations
of the advection of particle pairs in a synthetic turbulent velocity field
with prescribed Eulerian statistical features.

In first place, we deal with the probability distribution of
Lagrangian quantities in a self similar, Kolmogorov-like scaling flow
where we confirm the Richardson-Obukhov predictions. We then
investigate the effects of Eulerian intermittency on Lagrangian
statistics where we find ``Lagrangian intermittency'', i.e. deviations
of the scaling exponents from the Richardson values. These effects cannot
be captured by dimensional arguments alone. The simplest step beyond
dimensional considerations is to extend the multifractal description,
which is successfully used for Eulerian statistics, to Lagrangian
quantities. 
We will see that our numerical simulations confirm the Lagrangian 
multifractal predictions.

The intermittency corrections to relative dispersion are however 
rather small. Moreover, they can be hidden by the finite scaling
range. Huge Reynolds numbers are necessary in order to discriminate
clearly the scaling exponents. Within this framework,
we propose a new methodology for the analysis 
of relative dispersion data, which deals with Lagrangian statistics
at fixed spatial separation between particles. In particular,
the statistics of ``doubling times'' -- the time that two
particles spend while doubling their separation --
seems a very promising tool in data analysis.

In Section \ref{sec:2} we address the problem of relative dispersion
in fully developed, homogeneous and isotropic turbulence.
In Section \ref{sec:3} the construction of a self similar
synthetic turbulent field is described.
In Section \ref{sec:4} results on the Lagrangian statistics
of particle pairs  advected by a non-intermittent velocity field 
are presented.
In Section \ref{sec:5} the method of doubling times is introduced
and its advantages with respect to the usual fixed time statistics 
are discussed.
In Section \ref{sec:6} is shown how to build an intermittent velocity field,
and the effects of the Eulerian field on Lagrangian statistics are discussed.
In Section \ref{sec:7} conclusions are drawn.

\section{The Richardson law}
\label{sec:2}

We consider the dispersion of a pair of particles 
passively advected by an homogeneous, isotropic,
fully developed turbulent field. Due to the incompressibility of
the velocity field the particles will, on average, separate from each other
 \cite{Cocke69,Orszag70}. 
The statistics of pairs separation is conveniently summarized by
the probability density function of distances between couples
of particles at a given time, $p(\bbox{R},t)$, called {\em distance neighbor
function} by Richardson \cite{Richardson26}.

In view of the diffusive effect exerted by the turbulent motion on 
the advected particles, Richardson argued that the time evolution of the
distance neighbor function could be described by a proper 
diffusion equation
\begin{equation}
\frac{\partial p(\bbox{R},t)}{\partial t} =
\frac{\partial}{\partial R_j} \left(
K(R) \frac{\partial p(\bbox{R},t)}{\partial R_j} \right)
\label{eq:1.1}
\end{equation}
with a $R$ dependent scalar turbulent diffusivity $K(R)$.
From a collection of experimental data, Richardson was able to 
obtain his celebrated ``4/3'' law:
\begin{equation}
K(R) = \alpha R^{4/3}
\label{eq:1.1b}
\end{equation}
where $\alpha$ is a constant.
This choice for the diffusivity relied mainly on empirical grounds
(dependence of the vertical eddy diffusivity in the atmosphere 
with the altitude). The fortunate Richardson choice of a rational 
exponent demonstrates his faith in an underlying universal physical 
mechanism of simple form.

It is easy to realize that in three dimensions the solution of 
equation (\ref{eq:1.1}) is
\begin{equation}
p(\bbox{R},t)= N (\alpha t)^{-9/2}
\exp\left(-\frac{9 R^{2/3}}{4 \alpha t}\right)
\label{eq:1.1bis}
\end{equation}
where $N$ is a normalization factor, 
which immediately leads to the growth laws for the average separation
\begin{equation}
\langle R^{2n}(t) \rangle = \int dR R^{2n} p(\bbox{R},t) \sim t^{3n} \; .
\label{eq:1.2}
\end{equation}
A discussion of the validity of diffusive approximations
to the problem of pair dispersion in turbulence can be found
in chapter 24.4 of Monin-Yaglom \cite{MY75}. 

The scaling (\ref{eq:1.2}) can actually be derived by a simple dimensional 
argument due to Obukhov \cite{MY75} starting from Kolmogorov similarity law for 
velocity increments in fully developed turbulence
\begin{equation}
\langle | \delta \bbox{v}^{(E)}(\bbox{R}) | \rangle = 
\langle |\bbox{v}(\bbox{x}+\bbox{R})-\bbox{v}(\bbox{x})| \rangle \sim
R^{1/3}
\label{eq:2}
\end{equation}
with $R=|\bbox{R}|$.
The particle pair separates according to
\begin{equation}
{d \bbox{R} \over d t}= \delta \bbox{v}^{(L)}(\bbox{R})
\label{eq:3}
\end{equation}
where $\delta \bbox{v}^{(L)}$ represents the velocity difference 
evaluated along the Lagrangian trajectories.
Assuming $\delta v^{(L)}(R) \simeq |\delta \bbox{v}^{(E)}(R) |$ 
from (\ref{eq:2}) one
obtains $d R^2/dt \sim R \delta v^{(L)}_{\parallel}(R) \sim R^{4/3}$
and hence the Richardson law (\ref{eq:1.2}).
The assumption that the Lagrangian velocity difference has the same 
(Kolmogorov) scaling as the Eulerian one relies on the intuitive 
idea that the main contribution to
the rate of separation comes from eddies of a size comparable
to that of the separation itself.

After having outlined the classical arguments on relative dispersion,
two remarks are in order. 
First, that the determination of 
specific functional dependencies -- such as the shape of the 
distance-neighbor-function $p(\bbox{R},t)$ -- lies beyond 
the possibilities of similarity hypotheses, and thus calls for 
additional hypotheses, like equation (\ref{eq:1.1}).
Second, diffusive approximations have no possibility of capturing
intermittency effects on Lagrangian statistics. 
We have thus to face with supplementary hypotheses which have 
a considerable degree of arbitrariness, and whose content can be mainly
judged a posteriori. In this respect the use of synthetic velocity fields
represents a flexible framework for the study of the
detailed features of the statistics of pair dispersion. 

\section{The synthetic turbulent field}
\label{sec:3}

The generation of a synthetic turbulent field which reproduces 
the relevant statistical features of fully developed
turbulence is not an easy task. Indeed to obtain a physically 
sensible evolution for the velocity field one has to take into account 
the fact that each eddy is subject to the action of all
other eddies. Actually the overall effect amounts only 
to two main contributions, namely the 
sweeping exerted by larger eddies and 
the shearing due to eddies of comparable size.
This is indeed a substantial simplification, but nevertheless
the problem of properly mimicking the effect of sweeping
is still unsolved. 

It is relatively easy to construct a spatial, time
independent, self affine or multiaffine velocity field in any
dimension. In order to let the particle separate 
one has then to introduce some time dependence (at least in two
dimensions). This has been done in previous synthetic simulations
either by adding a large scale flow which sweeps the particles on
a quenched background of eddies \cite{EM96}, or by moving arbitrarily the
Eulerian structures \cite{FV98}. Of course, both the solutions are
rather unphysical. 

To get rid of these difficulties
we shall limit ourselves to the generation of a synthetic 
velocity field in {\em Quasi-Lagrangian} (QL) coordinates \cite{LPP97},
thus moving to a frame of reference 
attached to a particle of fluid $\bbox{r}_{1}(t)$.
This choice bypasses the problem of sweeping, since 
it allows to work only with relative velocities,
unaffected by advection. 
As a matter of fact there is a price to pay for the considerable 
advantage gained by discarding advection, and it is that 
only the problem of two-particle 
dispersion can be well managed within this framework.
The properties of single-particle Lagrangian statistics cannot,
on the contrary, be consistently treated.

The QL velocity differences 
are defined as 
\begin{equation}
\bbox{v}(\bbox{r},t) = \bbox{u}\left(\bbox{r}_{1}(t)+\bbox{r},t\right)
- \bbox{u}\left(\bbox{r}_{1}(t),t\right) \; ,
\label{eq:2.1}
\end{equation}
where the reference particle moves according to
\begin{equation}
\dot{\bbox{r}}_{1}(t) = \bbox{u}(\bbox{r}_{1}(t),t) \; .
\label{eq:2.2}
\end{equation}
These  velocity differences have the useful property
that their single-time statistics are the same as the Eulerian ones
whenever considering statistically stationary flows \cite{LPP97}.
For fully developed turbulent flows,
in the inertial interval of length scales where 
both viscosity and  forcing are negligible, the QL longitudinal 
velocity differences show the scaling behavior
\begin{equation}
\langle \left| \bbox{v}(\bbox{r}) \cdot 
\frac{\bbox{r}}{r} \right|^p \rangle \sim r^{\zeta_p}
\label{eq:2.3bis}
\end{equation}
where the exponent $\zeta_p$ is a convex function of $p$,
and $\zeta_3=1$. 
This scaling behavior is a distinctive 
statistical property of fully developed turbulent flows
that we shall reproduce by means of a synthetic velocity field.

In the QL reference frame the first particle is at rest in the origin
and the second particle is at
$\bbox{r}_{2}=\bbox{r}_{1}+\bbox{R}$, advected with respect to the
reference particle by the relative velocity
\begin{equation}
\bbox{v}(\bbox{R},t) = 
\bbox{u}\left(\bbox{r}_{1}(t)+\bbox{R},t\right) -
\bbox{u}\left(\bbox{r}_{1}(t),t\right)
\label{eq:2.3}
\end{equation}
By this change of coordinates 
the problem of pair dispersion in an Eulerian velocity field has 
been reduced to the problem of single particle dispersion in
the velocity difference field $\bbox{v}(\bbox{r},t)$.
This yields a substantial simplification:
it is indeed sufficient to build a velocity difference field 
with proper scaling features in the radial direction only,
that is along the line that joins the reference particle
$\bbox{r}_{1}(t)$ -- at rest in the origin of the QL coordinates --
to the second particle $\bbox{r}_{2}(t)=\bbox{r}_{1}(t)+\bbox{R}(t)$.
To appreciate this simplification, it must be noted that actually 
all moments of velocity differences
$\bbox{u}\left(\bbox{r}_{1}(t)+\bbox{r}',t\right) -
\bbox{u}\left(\bbox{r}_{1}(t)+\bbox{r},t\right)=
\bbox{v}(\bbox{r}',t)-\bbox{v}(\bbox{r},t)$ should display
power law scaling in 
$|\bbox{r}'-\bbox{r}|$. Actually these latter differences 
never appear in the dynamics
of pair separation, and so we can limit ourselves to fulfill
the weaker request (\ref{eq:2.3bis}). 
Needless to say, already for three particle dispersion one needs
a field with proper scaling in all directions.

We limit ourselves to the two-dimensional case, 
where we can introduce a stream function for the QL
velocity differences
\begin{equation}
\bbox{v}(\bbox{r},t) = \nabla \times \psi(\bbox{r},t) \; .
\label{eq:2.4}
\end{equation}
The extension to a three
dimensional velocity field is not difficult but more expensive
in terms of numerical resources.
Under isotropic conditions, the stream function 
can be decomposed in radial octaves as
\begin{equation}
\psi(\bbox{r},\theta,t) = \sum_{i=1}^{N}\sum_{j=1}^{n} 
            \frac{\phi_{i,j}(t)}{k_i} F(k_i r) G_{i,j}(\theta) \; .
\label{eq:2.5}
\end{equation}
where $k_i=2^{i}$.
Following a heuristic argument, one expects that at a given $r$
the stream function is essentially dominated by the contribution
from the $i$ term such that $r \sim 2^{-i}$.
This locality of contributions suggests a simple choice for 
the functional dependencies of the ``basis functions'': 
\begin{equation}
F(x) = x^2(1-x)\ \,\,\, \mbox{\rm for}\ \,\, 0\le x \le 1
\end{equation}
and zero otherwise,
\begin{equation}
G_{i,1}(\theta) = 1, \qquad G_{i,2}(\theta) = \cos(2 \theta + \varphi_i)
\end{equation}
and $G_{i,j}=0$ for $j>2$ ($\varphi_i$ is a quenched random phase).
It is worth remarking that this choice is rather general
because it can be derived from the lowest order expansion for small $r$
of a generic streamfunction in Quasi-Lagrangian coordinates.

It is easy to show that, under the usual locality conditions
for IR convergence, $\zeta_p <p$ \cite{RS78}, the leading contribution to
the $p$-th order longitudinal structure function $\langle |v_r(r)|^p \rangle$ 
stems from $M$-th term in the sum (\ref{eq:2.5}),
$\langle |v_r(r)|^p \rangle \sim \langle |\phi_{M,2}|^p \rangle$
with $r \simeq 2^{-M}$. 
If the $\phi_{i,j}(t)$ are stochastic processes with characteristic
times $\tau_i=2^{-2i/3}\,\tau_0$, zero mean and
$\langle |\phi_{i,j}|^p\rangle \sim k_i^{-\zeta_p}$,
the scaling (\ref{eq:2.3bis}) will be accomplished.
An efficient way of to generate $\phi_{i,j}$ is \cite{BBCCV98}:
\begin{equation}
\phi_{i,j}(t) = g_{i,j}(t)\, z_{1,j}(t)\,z_{2,j}(t)\cdots z_{i,j}(t)
\end{equation}
where the $z_{k,j}$ are independent, positive definite, identically
distributed random processes with characteristic time $\tau_k$, while
the $g_{i,j}$ are independent stochastic processes with zero mean,
$\langle g_{i,j}^2\rangle \sim k_i^{-2/3}$ and characteristic time
$\tau_i$. 
The scaling exponents $\zeta_p$ are determined by the probability
distribution of $z_{i,j}$ via
\begin{equation}
\label{eq:zetap}
 \zeta_p = \frac{p}{3} - \log_2\langle z^p\rangle \; .
\end{equation}
As a last remark we note that by simply fixing the $z_{i,j}=1$
we recover the Kolmogorov scaling.

In this Paper we shall consider either synthetic turbulent fields
without corrections to Kolmogorov scaling, i.e. $\zeta_p=p/3$, 
either fields whose
intermittency corrections to the Kolmogorov scaling, i.e., nonlinear
$\zeta_p$, are close to the experimental three dimensional turbulence
exponents \cite{Anselmet84}, see Table \ref{table1}.

\section{Lagrangian statistics in absence of Eulerian intermittency} 
\label{sec:4}

\begin{figure}[ht]
\epsfxsize=360pt\epsfysize=252pt\epsfbox{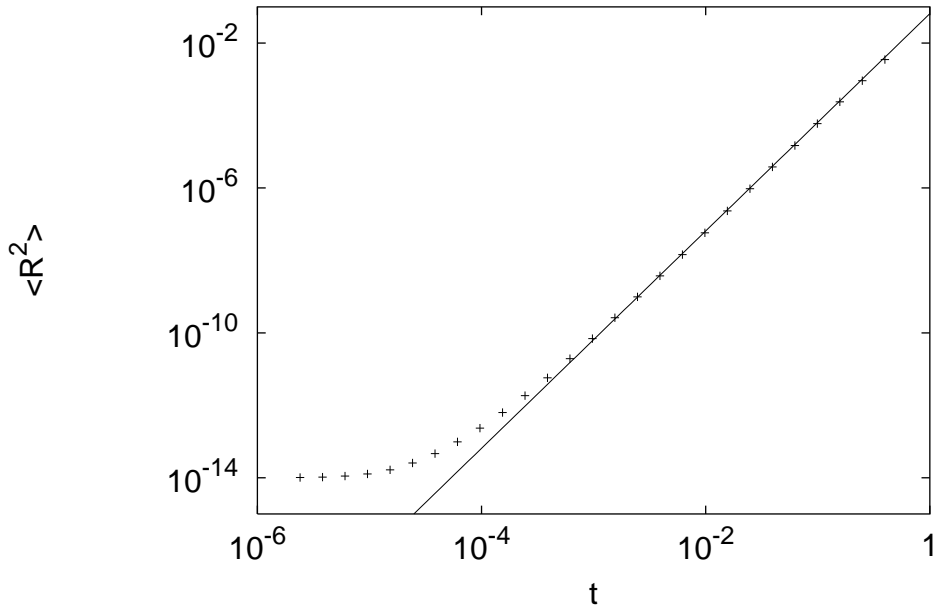}
\caption{Average variance of pair separation $\langle R^2(t) \rangle$ 
for simulation with $N=30$ octaves averaged over $10^{4}$ realizations.
The continuous line represent the Richardson scaling $t^{3}$.
}
\label{fig4.1}
\end{figure}

When the advecting velocity field has Kolmogorov scaling,
one expects Richardson law (\ref{eq:1.2}) to hold
for the variance of pair separation. 
This is indeed very well verified in our numerical simulations, 
as shown in Figure \ref{fig4.1},
over a range of separations of approximately 3 decades.
Observe that in the present example the effective Reynolds number,
defined as $Re = (k_{N}/k_{1})^{4/3}$ is already rather large,
$Re \simeq 10^{10}$. The scaling range in the relative dispersion
statistics is found to be strongly reduced with respect to that 
of structure functions.

\begin{figure}[ht]
\epsfxsize=360pt\epsfysize=252pt\epsfbox{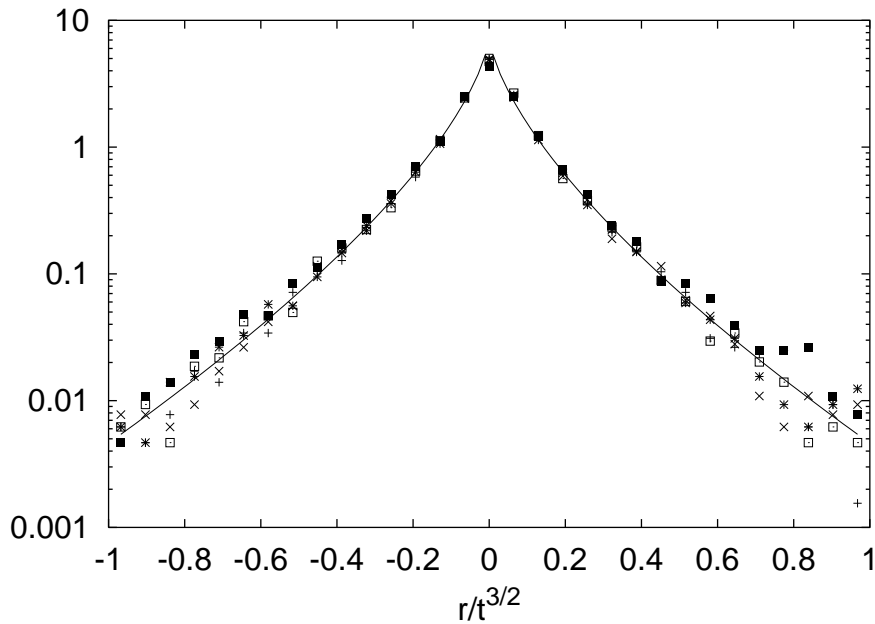}
\caption{Probability distribution functions of separations $R$
(distance neighbor function) rescaled with
the theoretical scaling $t^{3/2}$ for $10^{-3} < t < 0.25$ (i.e. in
the scaling region of previous figure). The continuous line represents
the Richardson distribution (\ref{eq:1.1bis}).
}
\label{fig4.2}
\end{figure}

Furthermore, by
similarity arguments, the distance neighbor function $p(\bbox{R},t)$
should assume the self similar, isotropic form (in two-dimensions)
\begin{equation}
p(\bbox{R},t)=Ct^{-3}\Phi(R/t^{3/2})
\label{eq:4.1}
\end{equation}
where $C$ is a normalization factor and $\Phi(\xi)$ is a universal function
whose shape is not predicted by similarity hypotheses.
We checked the validity of (\ref{eq:4.1}) by rescaling 
the numerically obtained distance neighbor functions with respect to
the theoretical average separation $<R> \sim t^{3/2}$.
The different rescaled pdf's, see Figure \ref{fig4.2}, collapse onto a 
unique curve, which represents the shape of the universal function 
 $\Phi(\xi)$ with $\xi=R/t^{3/2}$. 
The continuous line is the Richardson prediction
$\Phi(\xi) \sim \exp(-b \xi^{2/3})$ rephrasing eq. (\ref{eq:1.1bis}),
which is in good agreement with the data. 
This result supports the empirical picture that the relative diffusion
can be regarded as generated by a turbulent diffusivity,
according to (\ref{eq:1.1}).

\begin{figure}[ht]
\epsfxsize=360pt\epsfysize=252pt\epsfbox{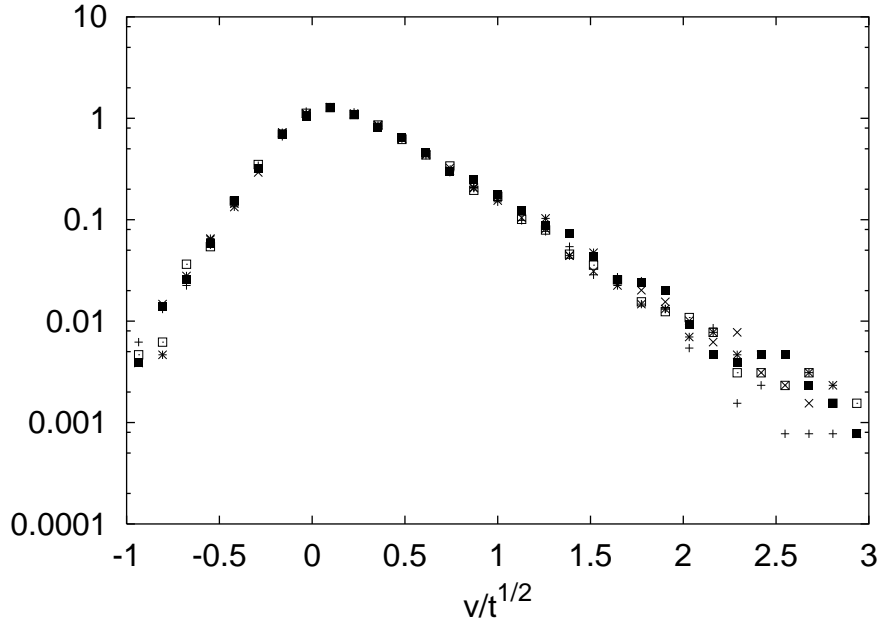}
\caption{Probability distribution functions of Lagrangian longitudinal
velocity differences at fixed $t$ rescaled with the theoretical scaling 
$t^{1/2}$ for the same values of $t$ as in figure \ref{fig4.2}.
}
\label{fig4.4}
\end{figure}

Another interesting statistics which can be investigated is the 
pdf of Lagrangian velocity differences along the trajectory,
$p_{L}(\delta v|t)$. Dimensional arguments give 
$\langle \delta v^2 \rangle \sim t$
showing the accelerating nature of Richardson dispersion \cite{MY75}. 
In Figure \ref{fig4.4} we plot the numerical computed pdf of 
$\delta v/t^{1/2}$. The clear collapse of the curves
for different times $t$ demonstrate the validity of the scaling
assumption.

Most of the previous works concerning the validation of the 
Richardson law have been focused mainly on the numerical prefactor 
(Richardson constant \cite{MY75}). 
We show that our results are realistic also for the 
Richardson constant $G_{\Delta}$
defined from the the pair dispersion law
 $R^2(t) = G_{\Delta}\,\overline{\epsilon}\, t^3 $.
The value of $\overline{\epsilon}$ can be obtained from the second order
Eulerian structure function, which reads
$S^{(E)}_2(R)= \langle |\delta v^{(E)}_{\parallel}(R)|^2 \rangle = 
       C_{L}\, \overline{\epsilon}^{2/3}\, R^{2/3}$
where $C_L$ is a universal constant related to the
Kolmogorov constant. According to the experimental measurements 
we fix $C_L=2.0$, leading to 
 $G_{\Delta} = 0.190 \pm 0.005$
for the Richardson constant which is in agreement with 
previous values \cite{MY75,EM96}.

It is worth remarking that, in the above expressions, $\overline{\epsilon}$
is {\em not} the average energy dissipation rate of the turbulent flow.
If it was so, we should be forced to conclude that in 
flows where no energy flux is present, no relative dispersion 
 takes place. 
Of course this is not the case, and the energy flux does not play a
significant role in determining the intensity of the turbulent diffusivity.
Strong turbulent diffusivity arises as a consequence of incompressibility
and non stationarity of the velocity field.
Thus $\overline{\epsilon}$ has thus to be intended as a dimensional
factor which simply rescales all the dimensional expressions
with respect to the large scale velocity $v_0$ and lenghtscale $\ell_0$.

\section{Doubling time statistics}
\label{sec:5}
\begin{figure}[ht]
\epsfxsize=360pt\epsfysize=252pt\epsfbox{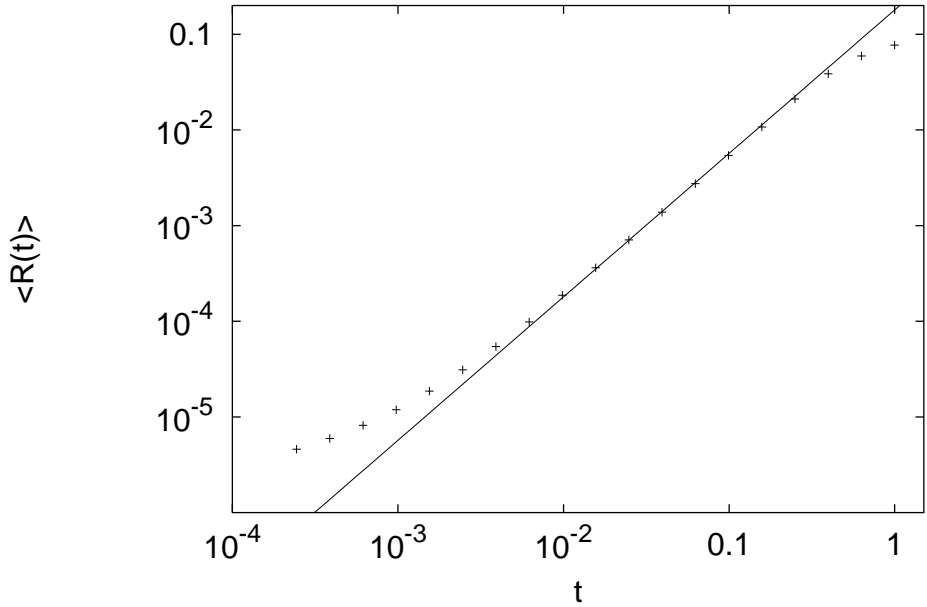}
\caption{Relative dispersion $\langle R(t) \rangle$ for $N=20$ octaves
simulation averaged over $10^4$ realizations. The line is the theoretical
Richardson scaling $t^{3/2}$.
}
\label{fig5.1}
\end{figure}

A closer look to Figure \ref{fig4.1} shows that the power-law scaling 
regime $\langle R^2(t) \rangle \sim t^{3}$ is observed only
well inside the inertial range.
To explain this effect let us consider a series of pair dispersion
experiments, in which a couple of particles is released at a 
separation $R_0$ at time $t=0$. At a fixed time $t_1$,
as customarily is done, we perform an average over all different
experiments to compute $\langle R^2(t_1) \rangle$.
But, unless $t_1$ is large enough that all particle pairs have 
``forgotten'' their initial conditions, our average will be biased.
This is the origin of the flattening of $\langle R^2(t) \rangle$
for small times, that we can call a crossover from initial 
condition to self similarity. In an analogous fashion
there is a crossover for large times, 
of the order of the integral time-scale, since some couples might have 
reached a separation larger than the integral scale, and thus 
diffuse normally, meanwhile other pairs still lie within the inertial
range, biasing the average and, again, flattening the curve
$\langle R^2(t) \rangle$.
This effect is particular evident for lower Reynolds numbers,
as shown in Figure \ref{fig5.1} for a simulation with $Re \simeq 10^{6}$.
This correction to a pure power law is far from being negligible for 
instance in experimental data where the inertial range is generally 
limited due to the Reynolds number and the experimental apparatus.
For example, references \cite{FV98,Fung92} show quite clearly the 
difficulties that may arise in numerical simulations with the standard
approach.

To overcome this difficulty
we propose an alternative approach which is 
based on the statistics at fixed scale, instead of at fixed time.
The method is in the spirit of a recently introduced generalization
of the Lyapunov exponent to finite size perturbation (Finite Size
Lyapunov Exponent) which has been 
successfully applied in the predictability problem \cite{ABCPV96} 
and in the diffusion problem \cite{ABCCV97}. 
Given a set of thresholds $R_n=R_0 2^{n}$ within the inertial range,
we compute the ``doubling time'' $T(R_n)$ defined as the time it takes 
for the particle separation to grow from one threshold $R_n$ to 
the next one $R_{n+1}$. 
Averages are then performed over many dispersion experiments.
The outstanding advantage of averaging at a fixed separation scale
is that it removes all crossover effects, since all sampled pairs
belong to the inertial range.

\begin{figure}[ht]
\epsfxsize=360pt\epsfysize=252pt\epsfbox{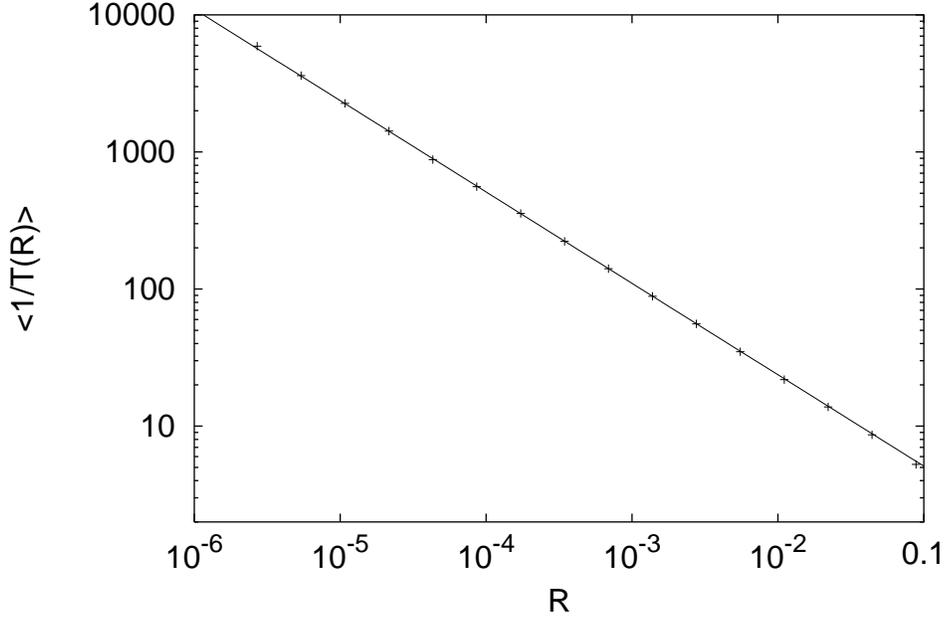}
\caption{Average inverse doubling time $\langle 1/T(R) \rangle$ for 
the same simulation of Figure \ref{fig5.1}. Observe the enhanced scaling
region. The line is the theoretical Richardson scaling 
$R^{-2/3}$ (\ref{eq:5.1}).
}
\label{fig5.2}
\end{figure}

The scaling properties of the doubling times is obtained by a
simple dimensional argument. The time it takes for particle separation
to grow from $R$ to $2 R$ can be estimate as $T(R) \sim R/\delta v(R)$;
we thus expect for the inverse doubling times the scaling
\begin{equation}
\left\langle {1 \over T^{p}(R)} \right\rangle \sim
       {\langle \delta v(R)^{p} \rangle \over R^{p}}
\sim R^{-2 p/3}
\label{eq:5.1}
\end{equation}
In  Figure \ref{fig5.2} the great enhancement in the scaling range 
achieved by using ``doubling times'' is well evident. 

The conclusion that can be drawn by this simple example is that
the doubling time statistics allows a much better estimation of
the scaling exponent with respect to the standard, fixed time,
statistics. This property will be used in the following section
for investigating the scaling law of the relative dispersion in
presence of Eulerian intermittency.

\section{The effect of Eulerian intermittency}
\label{sec:6}
In former literature there are very few attempts to investigate possible
corrections stemming from Eulerian intermittency 
\cite{Novikov89,CPV87,GP84,SWK87}. 
This is quite surprising compared with the enormous amount 
of literature concerning the intermittency correction for the Eulerian
statistics \cite{Frisch95,BJPV98}. 
This mismatch is partly due to the difficulty of having experimental 
checks of proposed theoretical corrections. 
The use of synthetic velocity fields provides a first benchmark which
is extremely easier and less expensive compared to experiments and direct 
numerical simulations.

\begin{figure}[ht]
\epsfxsize=360pt\epsfysize=252pt\epsfbox{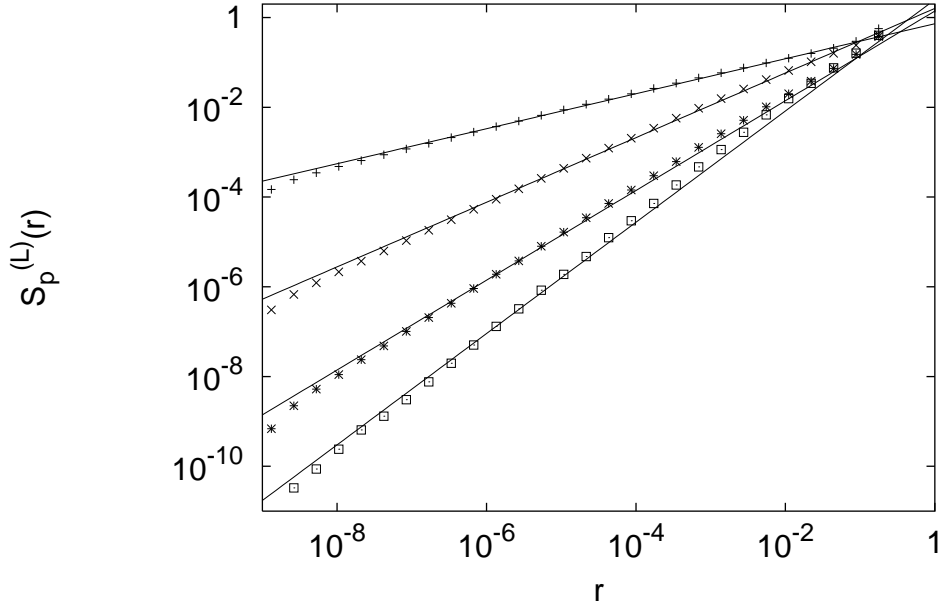}
\caption{Longitudinal Lagrangian structure functions $S^{(L)}_p(R)$ for 
$p=1,2,3,4$ (from top to bottom) for $N=30$ octaves intermittent 
velocity field. Average is over $10^{5}$ particle pairs.
The continuous lines represents the theoretical scaling
with exponents $\zeta_p$ given in Table \ref{table1}.
}
\label{fig6.1}
\end{figure}

In Figure \ref{fig6.1} we report the Lagrangian longitudinal structure 
functions 
$S^{(L)}_p(r)=\langle (\delta v^{(L)}_{\parallel}(r))^p \rangle$,
which are computed recording the Lagrangian velocity difference whenever
the pair separation equals $r$. 
Observe the wide inertial range over more than $10$ decades, 
corresponding to an integral Reynolds number $Re \simeq 10^{10}$.
Let us remark that, due to the average growth of particle separation,
also the first order Lagrangian structure function is non zero. 
The most interesting and non-trivial result is that scaling exponents for 
the Lagrangian structure functions show up to be almost exactly the same 
$\zeta_{p}$ of the Eulerian case (Table \ref{table1}).
In terms of the multifractal formalism \cite{Frisch95,PV87}, 
this result is restated by
saying that the fractal dimension $D(h)$ for the Lagrangian velocity
statistics is the same of the Eulerian one.

With this preliminary results in mind, we can extend the dimensional
argument for the Richardson law to the intermittent case by using
the multifractal representation.

From the definition
\begin{equation}
{d \over dt} \langle R^p \rangle = \langle R^{p-1} 
\delta v_{\parallel}^{(L)} \rangle 
\label{eq:6.1}
\end{equation}
by using the multifractal representation for the
velocity differences, we can write
\begin{equation}
{d \over dt} \langle R^p \rangle \sim
\int dh \, R^{p-1+h+3-D(h)} \, .
\label{eq:6.2}
\end{equation}
The time it takes for the pair separation to reach the scale $R$ 
is dominated by the largest time in the process and can be
dimensionally estimated as $t \sim R/\delta v \sim R^{1-h}$. 
Averaging over many realizations gives the expression
\begin{equation}
{d \over dt} \langle R^p \rangle \sim
\int dh \, t^{p+2+h-D(h) \over 1-h}
\label{eq:6.3}
\end{equation}
which, once evaluated by steepest descent method, gives the final result
$\langle R^p \rangle \sim  t^{\alpha_p}$ with scaling exponents
\begin{equation}
\alpha_p = \inf_{h} \left[{p+3-D(h) \over 1-h} \right]
\label{eq:6.4}
\end{equation}

In the case of intermittent velocity field, the relative dispersion
displays non linear scaling exponent $\alpha_p$ (see Table \ref{table1}).
However there is an interesting result, already obtained in 
\cite{Novikov89}, for the case $p=2$. From the general 
multifractal formalism one has that 
$3-D(h) \ge 1-3h$ and the equality is satisfied for the scaling 
exponent $h_3$ which realizes the third order structure function $\zeta_3=1$. 
From (\ref{eq:6.4}) follows that $\alpha_2=3$ and thus 
we have that the Richardson law $\langle R^2 \rangle \sim t^3$ 
is not affected
by intermittency corrections, while the other moments in general are.
We note that the previous argument leading to (\ref{eq:6.4}) is 
just one dimensional reasonable assumption which can be justified only
a posteriori by numerical simulations.
Other different assumptions are possible \cite{Novikov89,CPV87,GP84} leading
to different predictions. 

The scaling exponents satisfy the inequality
$\alpha_p/p < 3/2 $ for $p>2$: this amounts to say
that, as time goes by, the right tail of the pdf
of the separation $R(t)$ becomes less and less broad. 
In other words, due to 
the effect of Eulerian intermittency, particle pairs are more 
likely to stay close to each other than to experience a large
separation.

\begin{figure}[ht]
\epsfxsize=360pt\epsfysize=252pt\epsfbox{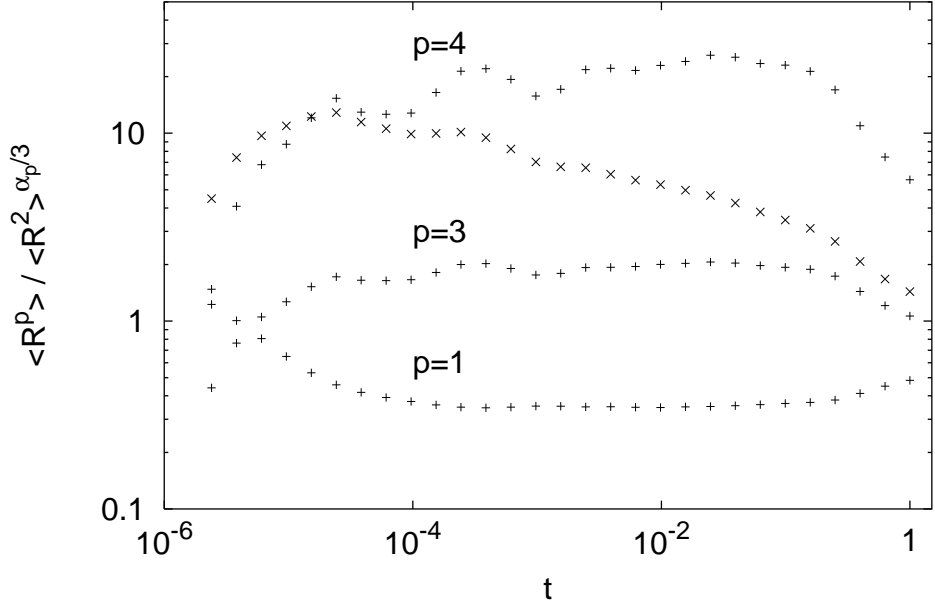}
\caption{
Relative dispersion $\langle R^p(t) \rangle$ rescaled with 
$\langle R^2(t) \rangle^{\alpha_p/3}$ for $p=1,3,4$ ($+$). The almost
constant plateau indicates a relative scaling in agreement with
prediction (\ref{eq:6.4}). For comparison we also plot 
$\langle R^4(t) \rangle$ rescaled with the non intermittent prediction
$\langle R^2(t) \rangle^2$ ($\times$) clearly indicating a 
deviation from normal scaling.
}
\label{fig6.2}
\end{figure}

In Figure \ref{fig6.2} we show the result of the computation of
$\langle R^p(t) \rangle$ for different $p$. 
We find that $\langle R^2(t) \rangle$ displays a clear $t^3$ scaling
law, but the scaling region becomes smaller for higher moments, 
making the determination of the exponents $\alpha_{p}$ rather difficult.
To overcome this difficulty we plot the moments $\langle R^p(t) \rangle$ 
compensated with $\langle R^2(t) \rangle^{\alpha_p/3}$ 
which should result constant according to (\ref{eq:6.4}).
For comparison we plot also the moment $p=4$ compensated assuming
normal scaling, i.e. $\langle R^4(t) \rangle \sim \langle R^2(t) \rangle^2$.
It is evident that prediction (\ref{eq:6.4}) is compatible with 
our numerical data, while the Richardson scaling (\ref{eq:1.2}) is not.
To be more quantitative, in Table \ref{table1}
we report the numerical $\alpha_p$ directly obtained by a fit of 
$\langle R^p(t) \rangle$. The numerical exponents, although 
affected by large uncertainty, are rather close to the theoretical ones.

The time doubling analysis discussed in Section \ref{sec:5}
reveals very useful in the case of Eulerian intermittency.
To see how the scaling of the doubling times are affected by
Eulerian intermittency,
we can give a dimensional estimate of the doubling time as
$T(R) \sim R/\delta v(R)$ and thus see that it fluctuates with the velocity
fluctuations. After averaging over many realizations we can write
\begin{equation}
\left\langle {1 \over T^{p}(R)} \right\rangle \sim
\int dh R^{p(h-1)} R^{3-D(h)} \simeq R^{\zeta_p-p} 
\label{eq:6.7}
\end{equation}
from which follows that the doubling time statistics contains the same
information on the Eulerian intermittency as the relative dispersion 
exponents (\ref{eq:6.4}). 
Let us remark that also in this case there 
is the exponent $\zeta_3-3=-2$ unaffected by Eulerian intermittency.

\begin{figure}[ht]
\epsfxsize=360pt\epsfysize=252pt\epsfbox{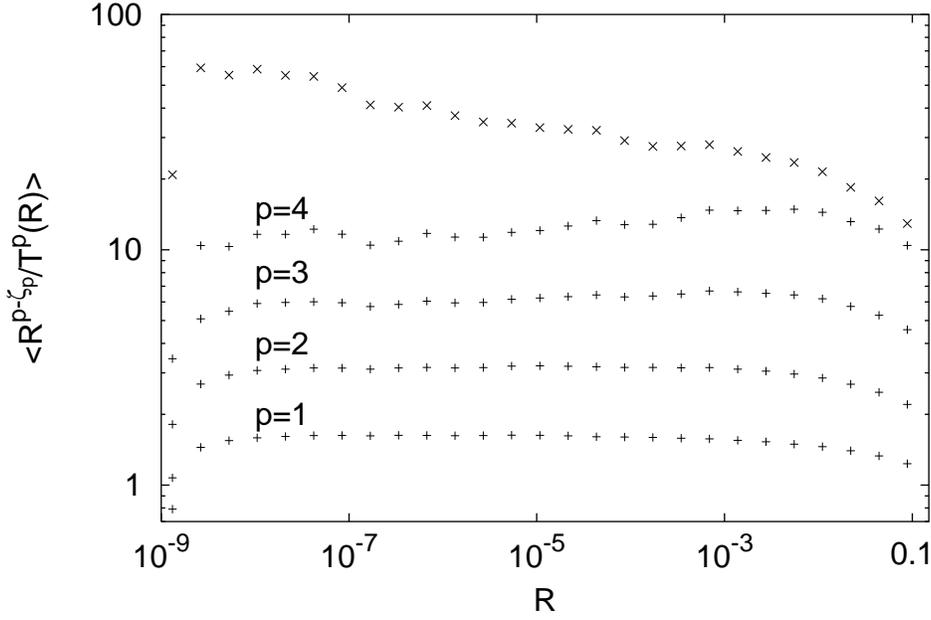}
\caption{
Inverse doubling times statistics $\langle 1/T^p(R) \rangle$
compensated with the multifractal prediction (\ref{eq:5.1}) 
$R^{\zeta_p-p}$ for $p=1,2,3,4$ ($+$). 
Inverse doubling times $\langle 1/T^4(R) \rangle$
compensated with the non intermittent prediction 
$R^{-8/3}$ ($\times$).
}
\label{fig6.3}
\end{figure}

As reported in Figure \ref{fig6.3} our prediction is very well 
verified in numerical simulations. The plot of the compensated
inverse time statistics clearly discriminates between multiaffine
scaling (\ref{eq:6.7}) and affine scaling (\ref{eq:5.1}) (here 
reported only for $p=4$).
Note that also in the present intermittent simulations, 
the scaling region for the inverse time statistics is wider 
than that of Figure \ref{fig6.2} and the scaling exponent can be
determined with much higher accuracy.
In Table \ref{table1} we report the theoretical exponent 
$\beta_p=\zeta_p-p$ of (\ref{eq:6.7})
compared with the direct numerical fit. The agreement is within
$2 \%$.

\subsection{Fluctuating characteristic times}
It must be pointed out that the construction of the synthetic field
here proposed shows an inconsistency with the expected statistical
properties imposed by Navier-Stokes equations.
Indeed, the characteristic time
the the eddies of size $R$ are chosen to be $\tau(R) \sim R^{2/3}$.
Actually, since by dimensional argument $\tau(R) \sim R/\delta v(R)$,
as long as $\delta v(R)$ are fluctuating quantities, the characteristic
times also fluctuate. This effect is definitely enhanced in presence
of intermittency, where the average characteristic times do not show
simple scaling \cite{LPP97}.
This finding could shed same doubt on the validity of the results 
obtained up to here.

We thus turn to a popular dynamical model of
turbulence which seems to have been developed exactly for our
purpose. Shell Models \cite{BJPV98} are deterministic models which 
displays a dynamic energy cascade. The model is built in terms of shell
variables $u_n$ which represent the velocity differences in a wavenumber
octave $k_n=k_0 2^n$ in Quasi-Lagrangian coordinates. With a suitable 
choice of parameters, the model develops chaotic dynamics which is 
responsible of intermittency correction to the Kolmogorov scaling 
exponents remarkably close to the experimental data \cite{BJPV98}.
Being a complete deterministic system, the Shell Model also displays
dynamic eddy turnover times with the same statistics expected 
for Navier-Stokes turbulence \cite{BBCT98}.

The particular model we use is a recently proposed Shell Model 
\cite{LPPPV98} for the complex variables $u_n$:
\begin{equation}
\frac{d u_n}{dt}=i k_n \left( u_{n+2}u_{n+1}^*
 -\frac{1}{4} u_{n+1}u_{n-1}^*+\frac{1}{8} u_{n-1}u_{n-2} \right)  
-\nu k_n^2  u_n +f_n
\label{eq:6.8}
\end{equation}
where $\nu$ is the viscosity and $f_n$ is a forcing term restricted
to the first two shells. 

\begin{figure}[ht]
\epsfxsize=360pt\epsfysize=252pt\epsfbox{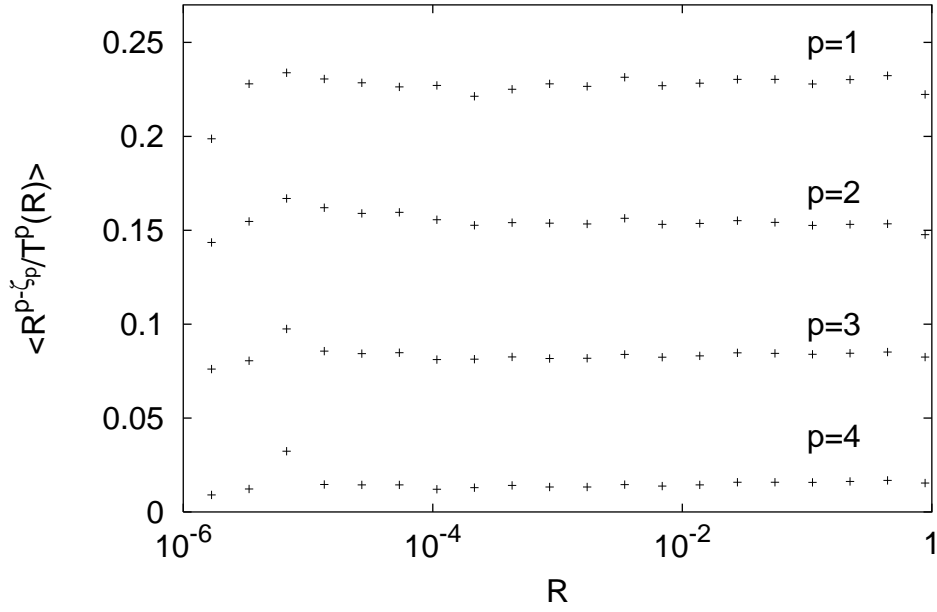}
\caption{
Inverse time statistics $\langle 1/T^p(R) \rangle$ compensated
with the multifractal prediction $R^{\zeta_p-p}$ for the Shell
Model simulation with $N=24$ shells, $\nu=10^{-8}$ and $k_0=0.05$. 
The average is over $10^4$ realizations of particle pairs.
}
\label{fig6.4}
\end{figure}

The QL stream function (\ref{eq:2.5}) can be written in terms of 
the $u_n$ variables by taking $\phi_{i,1}=\Re(u_i)$, 
$\phi_{i,2}=\Im(u_i)$. The scaling exponents for the Eulerian structure
functions $\zeta_p$ are numerically computed and listed in
Table \ref{table1}. 
In Figure \ref{fig6.4} we report the statistics of Lagrangian
doubling times compensated with the theoretical scaling (\ref{eq:6.7}). 
Also in this case there is evidence of anomalous scaling in agreement
with the multifractal prediction, confirming our previous finding 
with the stochastic velocity field. This result indicates that the
relative dispersion statistics is not very sensible to the details of
the time dependence of the Eulerian velocity field. We think this
is the main reason for which previous authors \cite{EM96,FV98} were
able to observe Richardson dispersion even with rather artificial
Eulerian dynamics.

\section{Conclusions}
\label{sec:7}

In this paper we have proposed a simple and efficient method for generating a 
time dependent, turbulent-like velocity difference field in Quasi-Lagrangian 
coordinates.
The synthetic flow is constructed with prescribed two-point scaling 
properties (structure functions) thus allowing extensive investigations
of the effects of intermittency on Lagrangian pair dispersion.

For non intermittent, Kolmogorov scaling, turbulence we find that
the original Richardson approach, based on a diffusion equation for
relative separation, in agreement with our simulations.
In the case of intermittent turbulence we have found that relative
dispersion displays anomalous scaling exponents and cannot be any
longer described as a self-similar process. 
Relative dispersion intermittency can be captured by a natural extension 
of the multifractal representation to Lagrangian quantities which leads
to the theoretical prediction of Lagrangian scaling exponents.

We have suggested a new approach based on the Lagrangian doubling
times which extends the scaling range with respect to the standard,
fixed time statistics, and is thus very promising for data analysis.

The present work are a first step towards the 
clarification of Lagrangian-Eulerian relationship in fully developed
turbulence. 
It would be extremely interesting to check our claims 
by mean of direct numerical simulations or laboratory experiments.

We thank L. Biferale for useful discussions.
This work has been partially supported by the INFM 
(Progetto di Ricerca Avanzata TURBO) and by the MURST
(program 9702265437).



\newpage

\begin{table}
\begin{tabular}{||c||c|c||c|c||c|c||c||} 
$p$ & $\zeta_p$ & $\zeta^{num}_p$ & $\alpha_p$ & $\alpha^{num}_p$ & 
$\zeta_p-p$ & $\beta^{num}_p$ & $\zeta^{SM}_p$ \\ \hline
1 & 0.390 & 0.39 & 1.59 & 1.56 & -0.610 & -0.62 & 0.39 \\
2 & 0.719 & 0.74 & 3.00 & 2.94 & -1.281 & -1.28 & 0.73 \\
3 & 1.0   & 1.04 & 4.32 & 4.27 & -2.0   & -1.99 & 1.01 \\
4 & 1.245 & 1.30 & 5.58 & 5.58 & -2.755 & -2.73 & 1.26 \\
5 & 1.461 & 1.54 & 6.80 & 6.88 & -3.539 & -3.49 & 1.49 \\
6 & 1.655 & 1.74 & 7.99 & 8.17 & -4.345 & -4.23 & 1.71 \\
\end{tabular}
\caption{ 
Theoretical and numerical fitted scaling exponent for the simulations
with intermittent velocity field. The number of shells is $N=30$
corresponding to an integral Reynolds number $Re \simeq 10^{10}$.
$\zeta_p$: Eulerian structure functions theoretical scaling exponents.
$\zeta^{num}_p$: Lagrangian structure function numerical scaling
exponents. 
$\alpha_p$: theoretical relative dispersion scaling exponents.
$\alpha^{num}_p$: numerical relative dispersion scaling exponents.
$\zeta_p-p$: theoretical doubling time scaling exponents.
$\beta^{num}_p$: numerical doubling time scaling exponents.
$\zeta^{SM}_p$: Eulerian scaling exponents for the Shell Model.
}
\label{table1}
\end{table}

\end{document}